\documentclass[twocolumn]{aastex631} 

\usepackage{ wasysym }

\usepackage[dvipsnames]{xcolor}
\usepackage{threeparttable}

\shorttitle{Have a Fit!}
\shortauthors{Rose et al.}

\begin{document}

\title{Modeling Stellar Collisions in Galactic Nuclei Using Hydrodynamic Simulations \\ and Machine Learning}

\author[0000-0003-0984-4456]{Sanaea C.\ Rose}
\affiliation{Center for Interdisciplinary Exploration and Research in Astrophysics (CIERA) and Department of Physics and Astronomy, Northwestern University, 2145 Sheridan Road, Evanston, IL 60201, USA }
\affiliation{NSF-Simons AI Institute for the Sky (SkAI), 172 E. Chestnut St., Chicago, IL 60611, USA}

\author[0000-0002-7444-7599]{James C.\ Lombardi, Jr.}
\affiliation{Department of Physics, Allegheny College, Meadville, Pennsylvania 16335, USA}

\author[0000-0002-0933-6438]{Elena Gonz\'{a}lez Prieto}
\affiliation{Center for Interdisciplinary Exploration and Research in Astrophysics (CIERA) and Department of Physics and Astronomy, Northwestern University, 2145 Sheridan Road, Evanston, IL 60201, USA }
\affiliation{NSF-Simons AI Institute for the Sky (SkAI), 172 E. Chestnut St., Chicago, IL 60611, USA}

\author[0000-0003-4412-2176]{Fulya K{\i}ro\u{g}lu}
\affiliation{Center for Interdisciplinary Exploration and Research in Astrophysics (CIERA) and Department of Physics and Astronomy, Northwestern University, 2145 Sheridan Road, Evanston, IL 60201, USA }
\affiliation{NSF-Simons AI Institute for the Sky (SkAI), 172 E. Chestnut St., Chicago, IL 60611, USA}


\author[0000-0002-7132-418X]{Frederic A. Rasio}
\affiliation{Center for Interdisciplinary Exploration and Research in Astrophysics (CIERA) and Department of Physics and Astronomy, Northwestern University, 2145 Sheridan Road, Evanston, IL 60201, USA }
\affiliation{NSF-Simons AI Institute for the Sky (SkAI), 172 E. Chestnut St., Chicago, IL 60611, USA}

\begin{abstract}

Nuclear star clusters can represent some of the most extreme collisional environments in the Universe. A nuclear star cluster like that of the Milky Way harbors a supermassive black hole at its center, which accelerates stars to high speeds ($\gtrsim 100$-$1000$ km/s) in a region where millions of other stars reside. Direct collisions occur in such high-density environments, where they can shape the stellar populations and influence the evolution of the cluster. 
We present a suite of a couple hundred
high-resolution smoothed-particle hydrodynamics (SPH) simulations of collisions between $1$~M$_\odot$ stars, at impact speeds representative of galactic nuclei.
We use our SPH dataset to develop physically-motivated fitting formulae for predicting collision outcomes. While collision-driven mass loss has been examined in detail in the literature, we present a new framework for understanding the effects of ``hit-and-run'' collisions on a star's trajectory.
We demonstrate that the change in stellar velocity follows the tidal-dissipation limit for grazing encounters, while the deflection angle is well-approximated by point-particle dynamics for periapses $\gtrsim0.3$ times the stellar radii.
We use our SPH dataset to test two machine learning (ML) algorithms, k-Nearest Neighbors and neural networks, for predicting collision outcomes.
We find that the neural network out-performs k-Nearest Neighbors and delivers results on par with and in some cases exceeding the accuracy of our fitting formulae. We conclude that both fitting formulae and ML have merits for modeling collisions in dense stellar environments, however ML may prove more effective as the parameter space of initial conditions expands.

\end{abstract}


\section{Introduction} \label{sec:intro}

The centers of galaxies can harbor both supermassive black holes (SMBH) and nuclear star clusters, with their coexistence depending on the mass and formation history of the galaxy \citep[e.g.,][]{Kormendy04,FerrareseFord05,Seth+08,Neumayer_Walcher_2012,ArcaSedda_CapuzzoDolcetta_2014,ArcaSedda_CapuzzoDolcetta2017,Nguyen+2018,Nguyen+19,Neumayer+20,Askar+21,Askar+22}. We focus on environments such as the center of the Milky Way Galaxy, which has a central $4 \times 10^6$~M$_\odot$ SMBH surrounded by a dense stellar cluster \citep[e.g.,][]{Ghez+05,Schodel+03,Gallego-Cano+18}.
These dense environments facilitate frequent close encounters between stars, including direct collisions \citep[e.g.,][]{BaileyDavies99,Rauch99,Dale2006,Dale+09,Davies+11,RubinLoeb,Mastrobuono-B,Rose+23}.

The collision rate in a nuclear star cluster can be estimated using an $n \sigma v$-type calculation, where $n$ is the number density of the colliders, or other stars, $\sigma$ is the cross-section of interaction, and $v$ is the relative speed. For direct collisions, the cross-section of interaction is the geometric cross-section plus some enhancement from gravitational focusing. We base this calculation on a Milky Way-like nuclear star cluster with a stellar density profile of the form $n \propto r^{-\alpha}$, normalized by $\rho_0 = 1.35 \times 10^6 \, M_\odot/{\rm pc}^3$ at $r_0 = 0.25 \, {\rm pc}$ \citep{Genzel+10rev}. We assume that $\alpha$ lies between $1.25$ and $1.75$ \citep[based on, e.g.,][]{BahcallWolf76,AharonPerets16,Gallego-Cano+18,LinialSari22}. For a cluster composed of $1$~M$_\odot$ stars surrounding a $4 \times 10^6$~M$_\odot$ SMBH, the collision timescale is $10$ Gyr at $0.1$ pc from the SMBH. Within this distance, the majority of Sun-like stars should experience at least one collision before evolving off the main-sequence \citep[see also][]{Rose+23,RoseMacLeod24}. About $10 \%$ of the stars, or $\sim 400,000$, reside within this collision-dominated region. However, even stars outside this region experience collisions. We estimate that a star at $0.5$ pc has about a $10\%$ chance of experiencing a direct collision over $10$~Gyr \citep[see also figure 1 in][]{Rose+23}. The overall rate of main-sequence stellar collisions is about $10^{-4}$ per year for a nuclear star cluster like the Milky Way’s \citep[consistent with, e.g.,][]{Freitag_Benz_2002confproc,AmaroSeoane23,Sidhu+25}.

One characteristic that sets nuclear star clusters apart from other collisional stellar systems, such as globular clusters, is that the SMBH dominates the gravitational potential in the inner $\sim 1$~pc.
Globular clusters have low velocity dispersions, and collisions there often result in stellar mergers \citep[e.g.,][]{Lombardi+96,Lombardi+02,Gonzalez+21,Rodriguez+22}. The SMBH, in contrast, accelerates stars on close-in orbits ($\lesssim 0.1$~pc) to high speeds \citep[e.g.,][]{Alexander99,Ghez+05,Genzel+03,Genzel+10rev}.
As a result, the impact speeds of colliding stars range from hundreds to thousands of km/s \citep[e.g.,][]{Rauch99,Rose+23}.
The most energetic collisions can drive significant mass loss from the stars, in some cases completely destroying them \citep[e.g.,][]{Lai+93,FreitagBenz,Balberg+13,Gibson+25,Ryu+24b,Dessart+24,Brutman+24}. 

Collisions have a number of implications for the stellar populations in galactic nuclei, astrophysical transients, and bulk properties of the cluster. For example, destructive collisions near the SMBH may be associated with transients both from the collisions themselves and from interactions of the ejected stellar material with the SMBH \citep[e.g.,][]{Balberg+13,AmaroSeoane23,Ryu+24b,Dessart+24,Brutman+24}. Destructive collisions
have also long been recognized as an important channel for shaping the stellar density profile in the inner region of a nuclear star cluster \citep[e.g.,][]{DuncanShapiro83,Murphy+91,David+87a,David+87b,Rauch99,Freitag+02,RoseMacLeod24,Balberg2024,AshkenazyBalberg24}. Furthermore, high-speed collisions and the peculiar stripped stars they produce have been connected to SMBH transients like tidal disruption events \citep[][]{Gibson+25,Rose&Mockler25}. 

The large range of speeds characteristic of galactic nuclei translates to diverse collision outcomes within the stellar population, depending on where the collision occurs in the cluster. While destructive collisions dominate in the inner $\sim 0.01$~pc, further from the SMBH, the velocity dispersion is lower and collisions can result in mergers 
\citep[e.g.,][]{Mastrobuono-B,Rose+23}.
These collisions can give rise to blue stragglers \citep[e.g.,][]{Lee&Nelson88,LeonarLinnell92,Lombardi+96,Lombardi+02,Sills+97,Sills+01} and represent a possible formation channel for the young-seeming massive stars observed in the inner $\sim 0.05$~pc of the Milky Way's Galactic center \citep[e.g.,][]{Ghez+03,Genzel+10,Alexander99,Rose+23}.

In addition to the relative speed between the stars, the outcome of a direct collision depends on the interplay of a number of initial conditions, including the impact parameter, mass ratio, and internal structures of the stars and therefore their ages. The effect of a collision on a star's mass and structure can only be fully understood using hydrodynamics simulations. Generally, simulations of collisional dynamics in nuclear star clusters
have needed to make physically-motivated assumptions about collision outcomes \citep[e.g.,][]{Mastrobuono-B,Rose+22,AshkenazyBalberg24,Rose&Mockler25}; over $10$~Gyr of evolution, tens of thousands of collisions may occur within the cluster, making it computationally prohibitive to run a hydrodynamics simulation to predict the outcome each time. Some nuclear star cluster models \citep[e.g.,][]{Rauch99,Rose+23} have leveraged fitting formulae developed from datasets of smoothed-particle hydrodynamics (SPH) simulations \citep[as provided by, e.g.,][]{Lai+93}. 
However, while outcomes like the mass loss from the stars and whether the collision results in a merger have been examined in detail, the effects of high-speed collisions on the trajectory of stars has been less well-studied \citep[for one of the few examples, see the appendix in][]{Rauch99}.

We present a new set of high-resolution SPH simulations of collisions between $1$~M$_\odot$ stars. We use our grid of $236$ simulations to build a physical intuition about the orbital effects of stellar collisions and develop fitting formulae to predict a star's post-collision velocity. We also re-visit fitting formulae for the mass loss and general collision outcome, e.g., if the collision destroys the stars, based on the literature. Furthermore, motivated by the vastness of the parameter space for direct collisions, which limits the effectiveness of even fitting formulae \citep[see discussion in][]{FreitagBenz}, we use our two-dimensional grid as a testing ground for machine learning (ML) techniques to model stellar collisions. Similar techniques have been successfully applied to collisions in planet formation simulations \citep[e.g.,][]{Winter+23,Crespi+24}, and recently, \citet{AmaroSeoane25} modeled the SPH dataset for stellar collisions from \citet{FreitagBenz} using gradient-boosted regression trees. This paper is organized as follows:

In Section~\ref{sec:SPH_methodology}, we describe our SPH simulations and methodology. Section~\ref{sec:grid_qualitative} provides a qualitative overview of the grid of collision outcomes. We then proceed to fitting formulae in Section~\ref{sec:fitting_formulae_section}. In Section~\ref{sec:capture_fit}, we describe the capture radius for a stellar merger to occur following a collision and find good agreement between our SPH dataset and the fitting formula from \citet{Lai+93}. Section~\ref{sec:f_ML_fit} focuses on the mass loss from the collisions. We develop fitting formulae for the effects of collisions on the stars' trajectories in Sections~\ref{sec:deflection_angle_fit} and \ref{sec:dv_v_fit}. In Section~\ref{sec:machinelearing}, we consider two different ML algorithms to model the SPH dataset: k-Nearest Neighbors and neural networks. We use classification in Section~\ref{sec:classification} to predict the category of collision outcome and regression in Section~\ref{sec:regression} to predict variables like the mass loss. Section~\ref{sec:ML_vs_FF} compares ML with our fitting formulae.  We conclude in Section~\ref{sec:conclusion}.

\section{Smoothed-particle Hydrodynamics Simulations} \label{sec:SPH_methodology}

We use the SPH code {\tt StarSmasher} \citep{Gaburov_GPU_SPH,Rasio_91_thesis} to simulate hyperbolic collisions between two $1$~M$_\odot$ stars. We obtain the structure of the star from a MESA profile at $2.5$~Gyr from \citet{Kiroglu+23}.  In our simulations, each star is represented with $N=10^5$ SPH particles that carry mass, position, velocity, and internal energy.
We verified that this resolution is sufficient by running test simulations at various particle numbers. We confirm that key quantities like mass loss are sufficiently converged by $N=10^5$ particles per star (see Appendix~\ref{sec:resolution_study}).
The local resolution is set by an adaptive smoothing length assigned to each particle. We employ a Wendland C4 smoothing kernel \citep{Wendland1995} for hydrodynamic interpolation and for gravity softening. To handle shocks and suppress unphysical particle interpenetration, an artificial viscosity is used together with a Balsara-type switch (see \citealt{Balsara1995,2015ApJ...806..135H}).
 
The particles are advanced using equations of motion obtained from a variational formulation that enforces proper evolution of energy and entropy \citep{Monaghan20002,2002MNRAS.333..649S,Lombardi2006}. 
Self-gravity in our calculations is evaluated by direct summation of particle-particle interactions on NVIDIA GPUs \citep{Gaburov_GPU_SPH}, consistent with energy-conserving formulations for particle-based gravity \citep{2007MNRAS.374.1347P}. Thermodynamics are modeled with an analytic equation of state that includes ideal gas and radiation pressure, as in \cite{Lombardi2006}: in particular, the pressure $P = \rho k T/\mu + aT^4/3$ and the specific internal energy $u = 3 kT/(2\mu) + aT^4/\rho$, where $\rho$ is density, $k$ is the Boltzmann constant, $\mu$ is the mean molecular mass, and $a$ is the radiation constant. Given $(\rho,u,\mu)$ for each particle, the temperature $T$ is obtained by solving the resulting quartic equation (implemented analytically as in \citealt{Lombardi2006}), after which the pressure $P$ is evaluated. For our current simulations of 1 $M_\odot$ main-sequence stars, radiation pressure contributes negligibly to the pressure in the parent stars but can become more important in shock heated regions.

After each impact, we evaluate the material gravitationally bound to each star. Our determination of the final bound masses follows the iterative procedure in \citet{Lombardi2006}. Guided by \citet{Nandez2014}, the boundness criterion employs mechanical energy, rather than the Bernoulli equation (i.e., specific enthalpy is excluded).
This choice avoids classifying shock-heated yet still gravitationally bound material as unbound (see \citealt{Nandez2014}, Sections~3.2 and~5.1). Accordingly, we run each simulation to a sufficiently late time that any unbound ejecta has expanded and cooled (rendering its internal energy dynamically unimportant), and the bound mass of each star has converged.
As our collisions never produce a stable binary as the ultimate end state, a common-envelope component need not be considered.  To determine the collision-induced velocity change and deflection angle, we propagate the relative two-body orbit from the final simulation snapshot to $r \to \infty$ and compute these quantities from the resulting asymptotic velocities of the two stars.

We explore a two dimensional parameter space of $v_{\infty}$ and $r_{p}$, where the former is the relative speed of the stars at infinity and the latter is the distance of closest approach between the stars. 
We model each encounter as an isolated two-body interaction in the center-of-mass frame and therefore do not include an external cluster potential or SMBH tidal field: the collision proceeds on stellar length scales and dynamical timescales, while any external potential varies on much larger scales. The main exception would be if the encounter results in a wide, weakly bound post-encounter pair that could be dissociated by SMBH tides (e.g., \citealt{Antonini+11}).\footnote{The largest apoapsis separation of any temporary binary formed in our simulations was $200$~R$_\odot$ ($\sim 1$~AU), which would need to be within $\sim 1 \times 10^{-3}$~pc of the SMBH to be tidally disrupted. This particular system had $r_p = 1.3$~R$_\odot$ and $v_\infty = 100$~km/s.}
Generally, we sample velocities at intervals of $200$ to $300$~km/s from $100$ to $5000$ km/s. We sample $r_{p}$ at intervals of $0.2$~R$_\odot$ from $r_{p} =0$ (head-on collisions) to $1.8$~R$_\odot$ (grazing collisions), with some exceptions. Most notably, we include four simulations with $v_{\infty} = 10$, $100$, $300$ and $600$ km/s for a close passage of $2.5$~R$_\odot$ to compare with expectations for tidal dissipation and tidal capture \citep[e.g.,][]{PressTeukolsky77,Lee&Ostriker86}. 
A list of input parameters and their ranges for the SPH simulations can be found in Table~\ref{tab:SPHinput}.

\section{Qualitative Results} \label{sec:grid_qualitative}

\begin{figure*}
        \centering
        
        \textbf{(A) Stellar merger:} $r_{\rm p}=0.2 R_\odot$, $v_{\infty}=900$ km\,s$^{-1}$~~~~~~~~~~~~~~~~~~~~~~~~~~~~~~~~~~~~~~~~~~~~~~~~~~~~~~~~~~~~~~~~~~~~~~~~~~~~~
        
        \includegraphics[trim={0 0 2.08in 0},clip,width=0.352\linewidth]{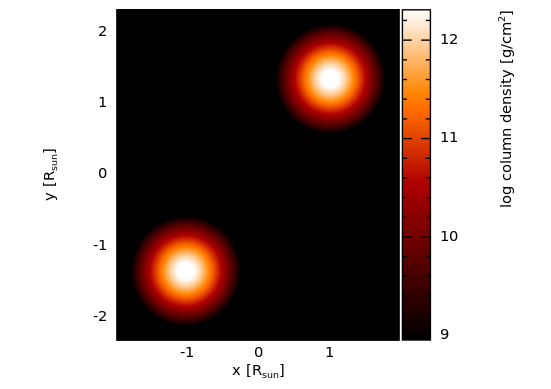}
        \includegraphics[trim={1.6in 0 2.08in 0},clip,width=0.251\linewidth]{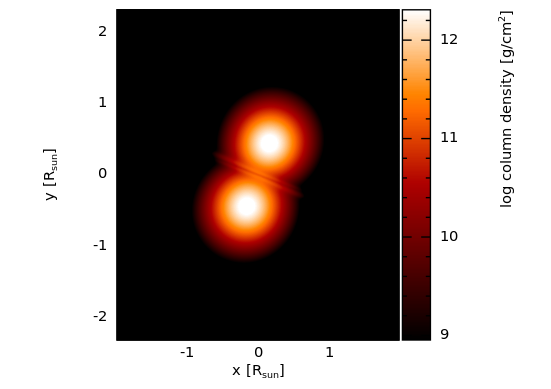}
        \includegraphics[trim={1.6in 0 0 0},clip,width=0.383\linewidth]{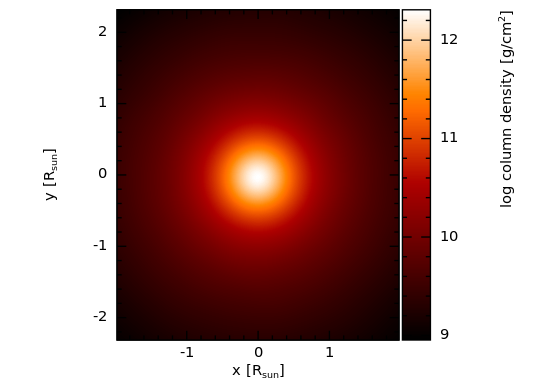} \\
~ \\
        
        \textbf{(B) Hit-and-run collision:} $r_{\rm p}=0.2 R_\odot$, $v_{\infty}=1900$ km\,s$^{-1}$~~~~~~~~~~~~~~~~~~~~~~~~~~~~~~~~~~~~~~~~~~~~~~~~~~~~~~~~~~~~~~~~

        \includegraphics[trim={0 0 2.08in 0},clip,width=0.352\linewidth]{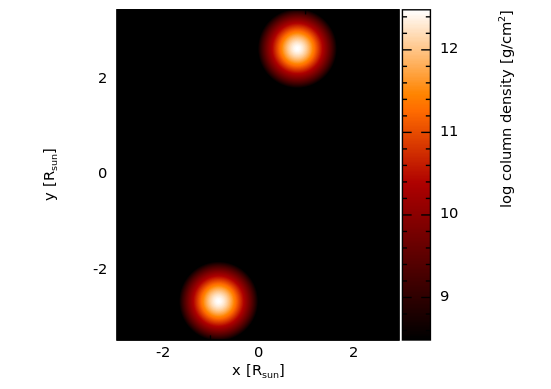}
        \includegraphics[trim={1.6in 0 2.08in 0},clip,width=0.251\linewidth]{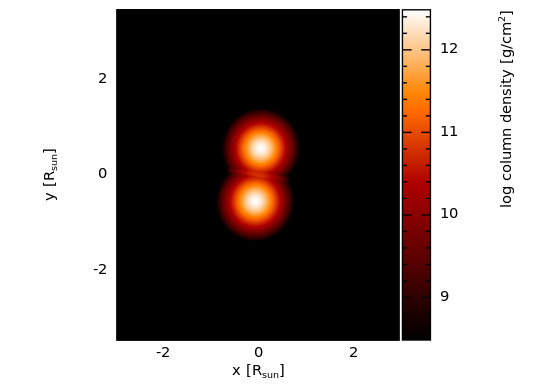}
        \includegraphics[trim={1.6in 0 0 0},clip,width=0.383\linewidth]{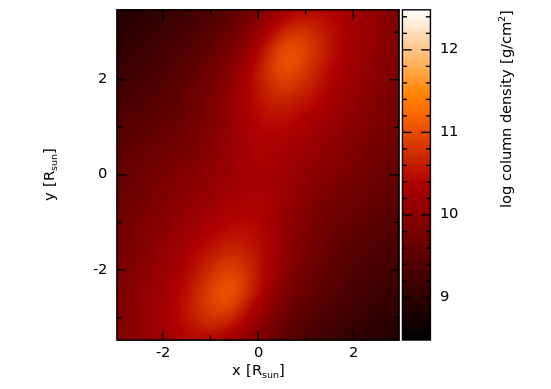} \\
~ \\

        \textbf{(C) Complete destruction:} $r_{\rm p}=0.2 R_\odot$, $v_{\infty}=3700$ km\,s$^{-1}$~~~~~~~~~~~~~~~~~~~~~~~~~~~~~~~~~~~~~~~~~~~~~~~~~~~~~~~~~~~~~~~~
    
        \includegraphics[trim={0 0 2.08in 0},clip,width=0.352\linewidth]{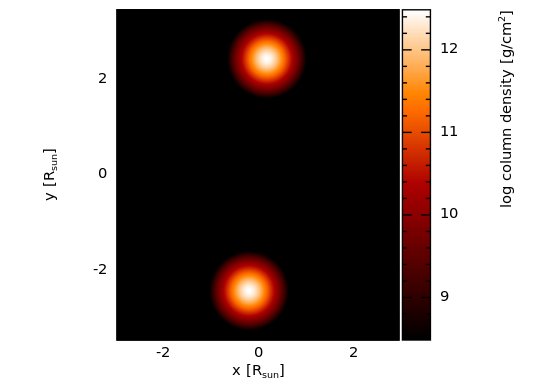}
        \includegraphics[trim={1.6in 0 2.08in 0},clip,width=0.251\linewidth]{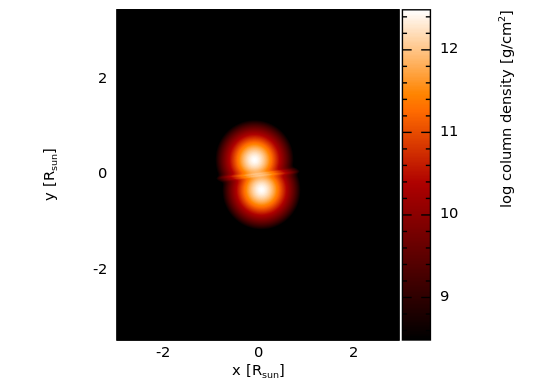}
        \includegraphics[trim={1.6in 0 0 0},clip,width=0.383\linewidth]{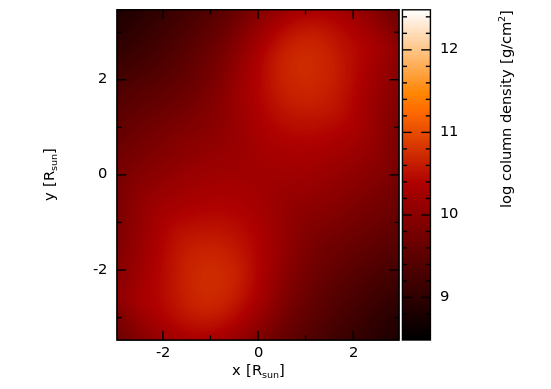} \\

        \caption{
        Various collisions of two identical $1\,M_{\odot}$ stars. \textbf{(A)} The top row illustrates the case with $r_{\rm p}=0.2 R_\odot$, $v_{\infty}=900$ km\,s$^{-1}$, resulting in a stellar merger. \textbf{(B)} The second row of snapshots shows $r_{\rm p}=0.2 R_\odot$, $v_{\infty}=3700$ km\,s$^{-1}$, in which both stars remain, having been stripped by the collisions. \textbf{(C)} The third row is for $r_{\rm p}=0.2 R_\odot$, $v_{\infty}=3700$ km\,s$^{-1}$: although two separate overdense regions are seen in the final panel, both dissipate as time progresses, leaving no final stars.  Color bars show column density on a logarithmic scale in g\,cm$^{-2}$. 
        }
        \label{fig:collision_summary}
    \end{figure*}

There are three possible qualitative outcomes for physical collisions between equal-mass stars:
\begin{enumerate}
    \item Stellar merger: if the relative speed or impact parameter between the stars is sufficiently low, the stars can capture one another and merge. We refer to this as a ``one star'' outcome. Figure~\ref{fig:collision_summary} shows snapshots from an example SPH simulation that resulted in a merger the first row.
    \item Hit-and-run collision: a subset of the collisions have too high a speed or too large an impact parameter to result in a merger. These collisions are best described as ``hit-and-run'' types of encounters; the stars physically collide, but they are not able to capture each other.
    We refer to this outcome as a ``two star'' case and show an example in the second row of Figure~\ref{fig:collision_summary}.
    \item Complete destruction: if the relative speed is sufficiently high and the impact parameter small, the high kinetic energy of the impact can effectively destroy the stars, a ``zero star'' outcome. Approaching this limit of total destruction, collisions drive large mass loss from the stars. We show an example of a zero-star outcome in the last row of Figure~\ref{fig:collision_summary}. 
\end{enumerate}

All of the examples in Figure~\ref{fig:collision_summary} have $r_p=0.2$~R$_\odot$. However, by increasing $v_\infty$, we achieve different collision outcomes. We note that in cases of unequal stellar masses, a collision can destroy one star while the other survives \citep{Lai+93,Gibson+25}, beyond the scope of this study.

\begin{figure*}
	\includegraphics[width=0.98\textwidth]{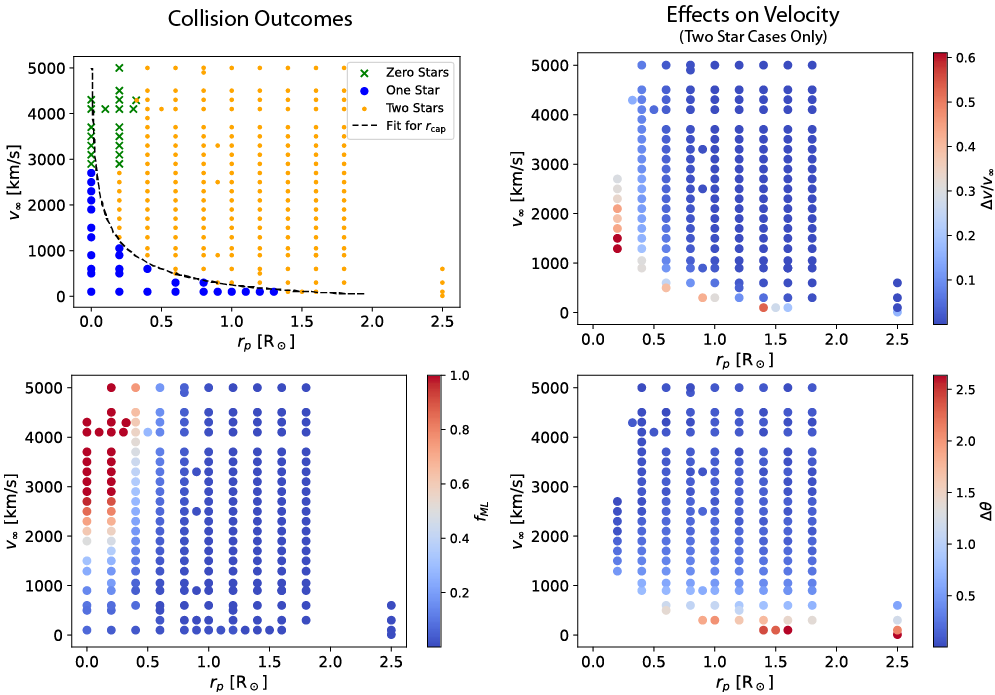}
	\caption{Parameters of interest from our grid of SPH simulations. \textbf{Upper left:} The general collision outcome for each pair of initial relative speed ($v_{\infty}$) and distance of closest approach between the stars ($r_{p}$) sampled. Blue circles represent cases where the collision culminated in a merger, i.e., resulted a single star. Green crosses indicate where the collision destroyed both stars. Orange dots represent cases where the speed was too high or the impact parameter too large for the stars to capture one another and merge. We show the fit for the boundary between mergers and higher-speed ``hit-and-run'' type collisions in the black dashed line (Eq.~(\ref{eq:capture_fit_ourunits}), adapted from \citet{Lai+93}'s equation 4). \textbf{Lower left:} The fractional mass loss from the stars ($f_{ML}$). A fractional mass loss of $1$ indicates a complete disruption of the stars. Fractional mass loss is maximized for small impact parameter and high speed. \textbf{Right column:} The effects of ``two star'' cases on the velocity vector of the stars. The upper panel shows the fractional decrease in speed ($\Delta v / v_{\infty}$) and the bottom panel, the deflection angle ($\Delta \theta$). The deflection angle is maximized for low-speed collisions, while the fractional change in speed is maximized for smaller $r_{p}$, corresponding to more physical overlap between the stars.}
     \label{fig:simulation_grids}
\end{figure*}

We show the results of our $236$ SPH simulations in Figure~\ref{fig:simulation_grids}. The upper left panel shows the general collision outcome for each simulation within our $v_{\infty}$-$r_{p}$ parameter space. The role of $v_\infty$ and $r_p$ in determining the collision outcome can be gleaned from the figure. For example, the entire upper right part of the parameter space comprises hit-and-run collisions, while fully destructive collisions only occur for low $r_p$ and high $v\infty$. We also plot a fitting formula (Eq.~\ref{eq:capture_fit_ourunits}) for the boundary between mergers and hit-and-run collision win black, described in more detail in Section~\ref{sec:capture_fit}.

The mass loss from the stars also characterizes the collision outcome. The transition from one and two star cases to zero star cases occurs when the fractional mass loss from the system nears $100 \%$. We show the mass loss from the system, expressed as a fraction $f_{\rm ML}$, for each of our simulations in the bottom left plot of Figure~\ref{fig:simulation_grids}. In equal-mass hit-and-run collisions, both stars will experience the same mass loss due to symmetry arguments. As such, $f_{ML}$ from the system is equal to the fractional mass loss from each star, $(M_{\star,i} - M_{\star,f})/M_{\star,i}$, where $M_{\star,i} = 1$~M$_\odot$. The mass loss is minimized for large $r_{peri}$ and low $v_{\infty}$.

In addition to the number of remnants and the mass loss from the system, we also examine the orbital effects of collisions. For mergers, the final star remains at rest in the center of mass frame; there is no asymmetric mass loss to impart a velocity kick. In simulations of dense stellar clusters, momentum conservation is used to treat collisions in this regime \citep[e.g.,][]{Rodriguez+22,Rose&Mockler25}. For zero star cases, the question of a star's final trajectory is rendered moot by its total destruction. We therefore focus our examination on the two star cases.

We approach changes to the star's velocity vector in two parts: magnitude and direction. We quantify the former using the fractional decrease in speed, $\Delta v/v_{\infty}$, while deflection angle $\Delta \theta$ 
quantifies the change in direction. 
We show the fractional change in speed and deflection angle in the right column of Figure~\ref{fig:simulation_grids}. The change in speed is maximized for small $r_{p}$ and, to a lesser degree, small $v_{\infty}$. In other words, collisions with the most physical overlap between the stars have the strongest effect on the speed, while for large $v_{\infty}$, the stars have very little time to act on one another. Similarly, the deflection angle is largest for small $v_{\infty}$ because it is easier to deflect slower moving stars. Generally, the relationship between fractional mass loss and $v_{\infty}$ trends in the opposite direction as the velocity changes. Therefore, most collisions that lead to high $f_{\rm ML}$ tend not to affect the trajectory of the stars significantly.

\begin{table*}
\centering
\caption{Input parameters for SPH simulations.}
\small
\begin{tabular}{l l c c c c c c}
\hline
\textbf{Interaction Type} & \textbf{Total Number} & \textbf{Mass} & \textbf{Radius} & \textbf{Stellar Age} &  \textbf{$r_{p}$} & \textbf{$v_\infty$}   \\
\hline
Direct Collision  & $232$ & $1$~M$_\odot$ & $1.05$~R$_\odot$ & $2.5$ Gyr & $0$-$1.8$~R$_\odot$ & $100$-$5100$~km/s  \\
Close Passage  & $4$ & $1$~M$_\odot$ & $1.05$~R$_\odot$  & $2.5$ Gyr & $2.5$~R$_\odot$ & $100$-$600$~km/s  \\
\hline
\end{tabular}
\begin{tablenotes}
\small
 \item All collisions are equal mass. Stellar profile was generated in MESA.
\end{tablenotes}
\label{tab:SPHinput}
\end{table*}

\section{Fitting Formulae} \label{sec:fitting_formulae_section}

To complement our qualitative understanding of collisions as described above, we develop physically-motivated fitting formulae to quantify their effects on the stars.

\subsection{Capture Boundary} \label{sec:capture_fit}

For two stars to merge during a close interaction, they must dissipate their relative kinetic energy, $\propto v_\infty^2$, to become gravitationally bound. Stars have two possible mechanisms for dissipating kinetic energy during close interactions: tidal dissipation and shock dissipation. The first mechanism, tidal dissipation, is sufficient for the stars to capture each other only in the regime where $v_\infty \ll v_{esc}$ and can produce bound systems \citep[e.g.,][]{Fabian+1975,PressTeukolsky77,Lee&Ostriker86,Ray+tidal_87,Kochanek92,Mardling95,Mardling95b}. As the relative speed between the stars increases, the system reaches a critical point where the stars must pass within a periapsis distance $r_p$ less than the sum of their radii to become bound, necessitating a physical collision \citep[e.g.,][]{Lee&Nelson88,BenzHills92}. As $v_{\infty}$ increases further, the distance of closest approach needed for the stars to become bound continues to decrease, and shocks
become the dominant mechanism to dissipate the kinetic energy \citep[e.g.,][]{Lai+93}.

We define the capture radius as the critical distance of closest approach needed for the stars to become gravitationally bound. We find that the fit for the capture radius from \citet[][see their equation 4]{Lai+93} performs well in light of our new SPH data. We note that \citet{Lai+93} fit in terms of a dimensionless velocity parameter $v_\infty^\prime = v_\infty/\sqrt{GM_\star/R_\star}$, while in this work, we typically fit in terms of the quantity $v_\infty/v_{esc}$, where $v_{esc} = \sqrt{2 G M_\star/R_\star}$ is the escape speed from the surface of the star. Re-writing their equation to be consistent with our notation, we arrive at the general form:
\begin{eqnarray} \label{eq:capture_fit_ourunits}
\frac{r_\mathrm{cap}}{R_{1}+R_{2}} = A_\mathrm{cap} \left( \frac{0.112}{v_\infty/v_{esc}} \right)^\eta,
\end{eqnarray}
where $A_{\rm cap}$ is a dimensionless constant, $\eta = 0.18+\sqrt{v_\infty/(5.65v_{esc})}$, and $R_{1}+R_{2}$ is the sum of the radii of the two colliding stars. We remind the reader that in this work $R_{1}=R_{2} = R_\odot$. As $v_\infty$ becomes much less than $v_{esc}$, $\eta$ approaches $0.18$. Thus, Eq.~\ref{eq:capture_fit_ourunits} reproduces the tidal capture limit for $n=3$ polytropes at small $v_\infty$  \citep[e.g.,][]{PressTeukolsky77,Lee&Ostriker86,Lai+93}.

\citet{Lai+93} fit $A_{\rm cap}$ for different values of a dimensionless parameter $\alpha$, which they define to be $\propto M_\star$. While they only provide values of $A_{\rm cap}$ for $\alpha = 1$ and $10$, corresponding to $M = 5$ and $50$~M$_\odot$, we extrapolate the value of $A_{\rm cap}$ for $1$~M$_\odot$ stars. Assuming a power-law fit to the values for equal-mass collisions in \citet{Lai+93} table $1$, we find $A_{cap} \propto \alpha^{0.064}$, giving the relation $A_{cap} = 0.95 \times \left(\frac{M_\star}{5M_\odot}\right)^{0.064}$. Consequently, $A_\mathrm{cap} = 0.857$ for $1$~M$_\odot$ stars. We plot the capture radius as a function of $v_\infty$ in the upper left panel of Figure~\ref{fig:simulation_grids} in black.

All physical collisions that result in a bound system ultimately merge the stars. By construction, the initial $r_p$ ensures physical contact between the stars. Even if the stars do not promptly merge on the first impact and instead form a binary, they can only return to periapsis and collide again. Dissipation works to tighten the system, and any binary configuration is temporary.
Within our parameter space of physical collisions, we had a couple of systems with low $v_\infty$ that made a few passes before merging (e.g., $r_p = 1.2$~R$_\odot$ and $v_\infty = 100$~km/s).
This outcome, that systems undergoing multiple passages following a physical collision ultimately coalesce, has been recognized in previous hydrodynamic studies \citep[e.g.,][]{BenzHills87,BenzHills92,Lai+93}. While we focus on the outcomes of physical collisions in this paper, even tidal capture from close passages (non-contact encounters with $r_p>2$~R$_\odot$) can lead to mass transfer and/or mergers \citep[e.g.,][]{Ray+tidal_87,Lee&Nelson88,Kochanek92,Podsiadlowski96,FaberRasioWillems05,MacLeod+22_tidal}. The deposition of energy in the stellar structure can even drive the expansion of the stars such that they physically collide \citep[e.g.,][]{BenzHills87,Lai+93}.

\subsection{Fractional Mass Loss} \label{sec:f_ML_fit}

\begin{figure*}
\begin{center}
	\includegraphics[width=0.9\textwidth]{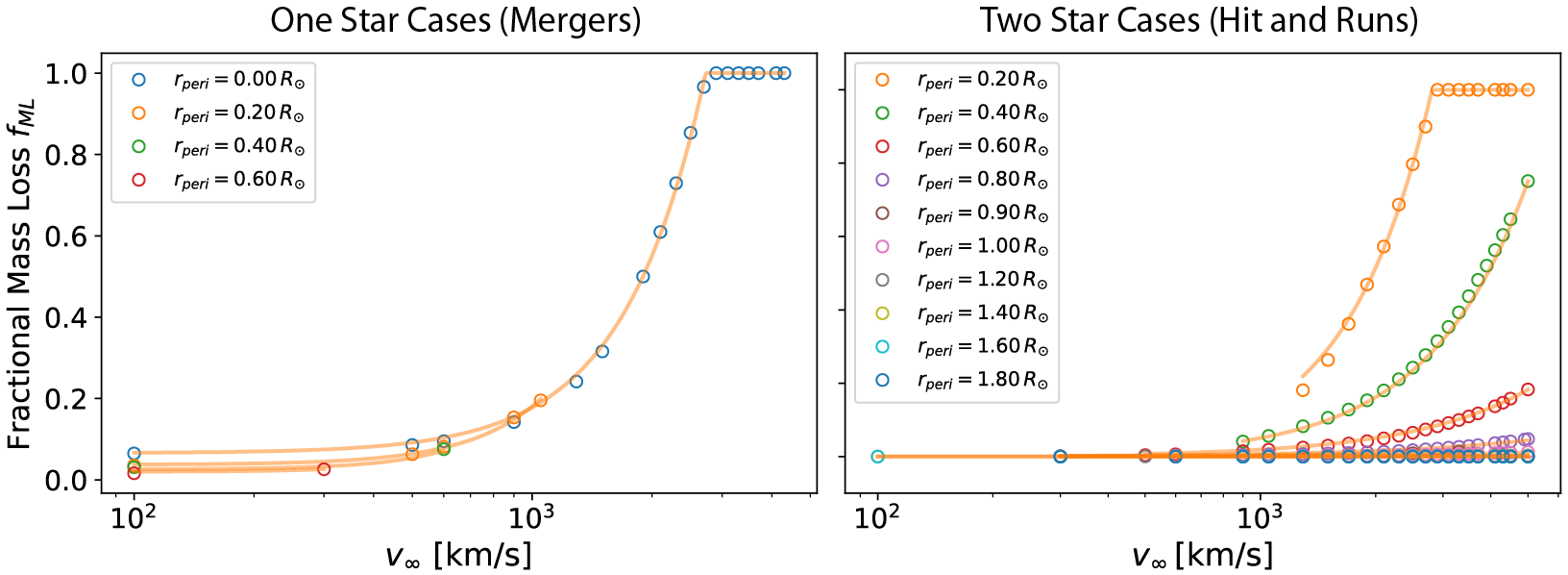}
	\caption{Fractional mass loss ($f_{ML}$) from the SPH simulations versus $v_\infty$ for different values of $r_p$. We divide the data up into mergers, or ``one star'', cases and hit-and-run type collisions, or ``two star'' cases. However, we also include relevant ``zero star'' cases -- total destruction, $f_{ML} = 1$ -- to show the transition between the other two classes and fully destructive collisions. The orange lines represent our fits from Eq.~\ref{eq:fML_mergers} (left) and  Eq.~\ref{eq:fML_highspeed} (right), which reproduce the transitions to $f_{ML} = 1$.}
     \label{fig:fractional_mass_loss_fits}
\end{center}
\end{figure*}

The framework for understanding collision-driven mass loss from \citet{Lai+93} informs our initial approach to the fitting for this parameter. As described in their section $2.2$, the mass ejected by a stellar collision should depend on the ratio of the kinetic energy, $\mu v_{\infty}^2$ where $\mu$ is the reduced mass, to the binding energy of the stars. Furthermore, hyperbolic collisions eject shock-heated gas in quantities dependent on the stellar structure. For a head-on collision, the fractional mass loss from the system should roughly be $A+B \times v_\infty^2$. 

As described in Section~\ref{sec:grid_qualitative}, we quantify the mass loss using the fraction $f_{\rm ML} = \frac{M_f-M_i}{M_i}$, where $M_f$ ($M_i$) is the final (initial) mass of the system. For collisions of equal-mass $1$~M$_\odot$ stars, symmetry arguments demand that the total mass loss from the system is equivalent to the mass loss from each star, $f_{\rm ML} = \frac{M_{f,\star}-M_\odot}{M_\odot}$. We begin from a fit with the form $f_{\rm ML} = A+B \times v_\infty^a$. In a departure from previous studies in the literature \citep[][]{Lai+93,Rauch99}, we fit the fractional mass loss for the one and two-star cases separately. While we focus on equal-mass collisions, future work may expand the fitting formulae to collisions with mass ratio $q \neq 1$. The mass loss from a merger product in the $q \neq 1$ regime can still be expressed as a single quantity. However, in hit-and-run collisions, stars of different masses will experience different degrees of mass loss, and the fit will need to be expanded to encompass the relationship between the mass loss fraction from the primary star and that of the secondary star.

For the merger regime in our simulation grid, we find that the fractional mass loss can be fit by: 
\begin{eqnarray} \label{eq:fML_mergers}
f_{\rm ML} =  \frac{0.0658}{1 + 3.85 \frac{r_{p}}{R_\odot}} + 0.0425 \left(\frac{v_{\infty}}{v_{\rm esc}} \right)^{2.08+2.19 \frac{r_{p}}{R_\odot}}.
\end{eqnarray}
Here, we have normalized $v_\infty$ by the escape speed from the surface of the star, $\sqrt{2GM_\odot/R_\odot}$, a form which reflects the ratio of the kinetic energy of the collision to the binding energy of the stars. While we have normalized the periapsis distance by the radius of the star, it is trivial to re-write the equation in a more generalized form that depends on $r_p/(R_{1}+R_{2})$, or $2 R_\odot$ for our system. The first term in Eq.~\ref{eq:fML_mergers} is the fit for parabolic collisions from \citet{Lombardi+02}. Our fit therefore reproduces the expectation in the regime of $v_\infty \ll v_{esc}$. We note that since Eq.~\ref{eq:fML_mergers} can give values greater than $1$, we take the maximum between $1$ and Eq.~\ref{eq:fML_mergers} as the true mass loss.

The best fit for the two-star cases, or hit-and-run collisions, is:
\begin{eqnarray} \label{eq:fML_highspeed}
f_{ML} =  0.12 \, e^{-4.17 \frac{r_p}{R_\odot}} \times \left(\frac{v_{\infty}}{v_{esc}} \right)^{2.23-1.39 \frac{r_{p}}{R_\odot}}.
\end{eqnarray}
The fractional mass loss is maximized for close to head-on collisions. To capture this behavior, we make the coefficient in the fit decay exponentially. As is the case for the merger fitting formula, the fractional mass loss is given by the maximum between Eq.~\ref{eq:fML_highspeed} and $1$.


We show our fitting formulae for mergers and hit-and-run collisions versus the SPH data in the left and right panels of Figure~\ref{fig:fractional_mass_loss_fits}, respectively. The data points are colored by $r_p$, while the orange curves show the fitting formulae predictions for the fractional mass loss. Both Eq.~\ref{eq:fML_mergers} and Eq.~\ref{eq:fML_highspeed} are accurate to within $\sim 4\%$ percent for all of our data. Furthermore, we reproduce the transition from surviving stars to total destruction ($f_{\rm ML} = 1$) for the head-on case ($r_p = 0$) in the left panel. Collisions with $r_p > 0$ must first transition from mergers to two-star collisions, shown on the right, before reaching the threshold for full destruction. It is therefore Eq.~\ref{eq:fML_highspeed}, the fitting formula for two-star collisions, that reproduces the transition to fully destructive collisions in the regime where $r_p>0$. 

\subsection{Deflection Angle} \label{sec:deflection_angle_fit}


\begin{figure*}
\begin{center}
	\includegraphics[width=0.99\textwidth]{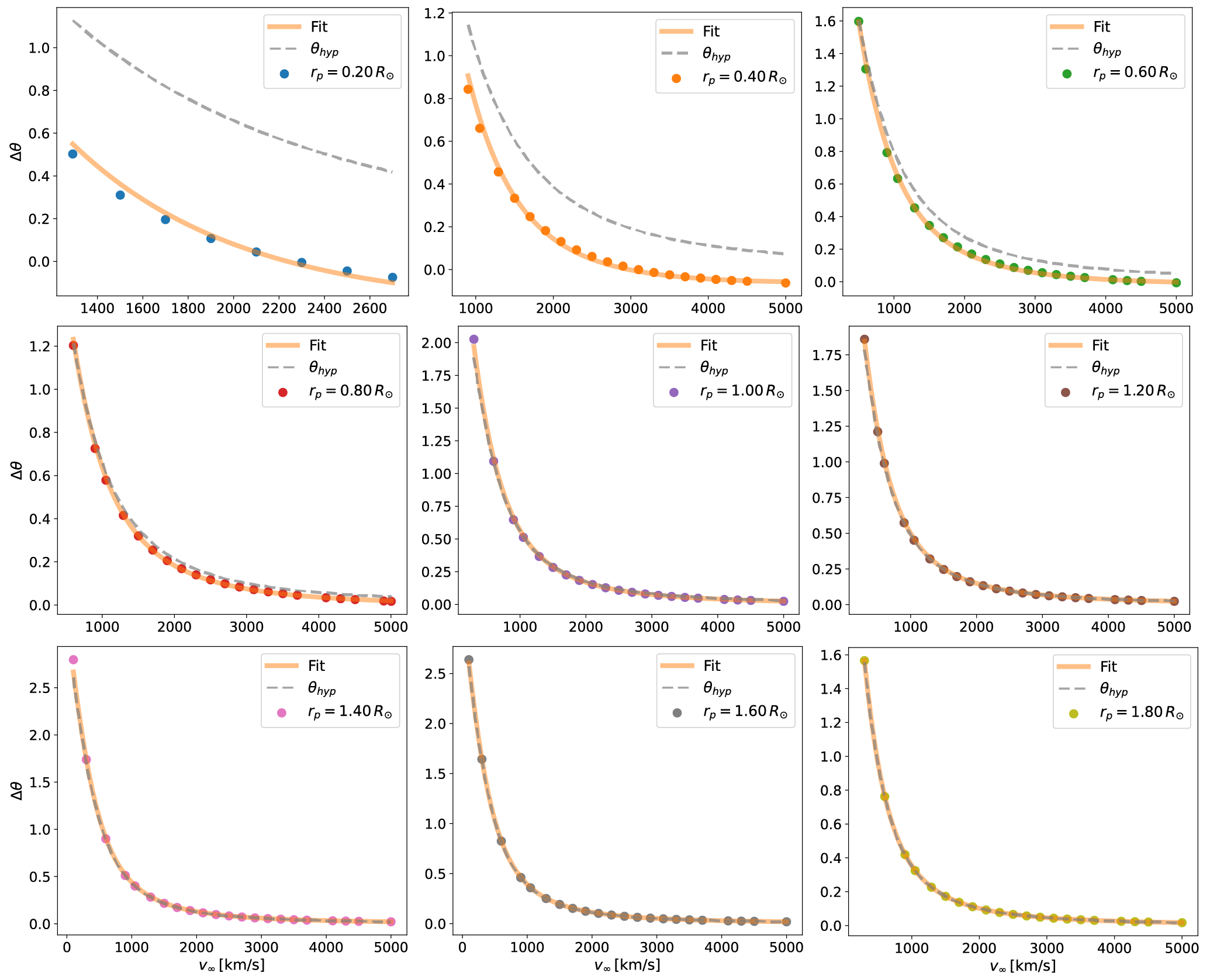}
	\caption{Deflection angle following a hit-and-run (two-star) collision versus $v_\infty$ for all of our values of periapsis with more than $5$ data points, which are the SPH data used to tailor the fitting formula given by Eq.~\ref{eq:theta_fit}. Dots represent the SPH results, while the grey dashed curve is the deflection angle for a hyperbolic encounter between point particles (Eq.~\ref{eq:theta}). The orange solid curves represent the predictions of our fitting formula (Eq.~\ref{eq:theta_fit}), which is generally accurate to within $4$ degrees for all of our data. We note that for $r_p>0.6$~R$_\odot$, the equation for a purely hyperbolic encounter predicts the deflection angle to within $0.1$ radians, or $6$ degrees.}
     \label{fig:deflection_angle}
\end{center}
\end{figure*}

Hyperbolic encounters between point particles provide an analytical framework within which we can understand two-star collisions.
In a purely hyperbolic encounter without a physical collision, stars will be deflected by angle $\theta_{\rm hyp}$ in their center of mass frame, though the speed of the stars at infinity will be unchanged due to energy conservation. The deflection angle can be written as:
\begin{eqnarray} \label{eq:theta}
\theta_{\rm hyp} = 2 \arctan(b_{90}/b) \ ,
\end{eqnarray}
where $b$ is the impact parameter and $b_{90}$ is the impact parameter needed for a $90 ^{\circ}$ deflection. $b_{90}$ equals $ G(2 M_\odot)/v_{\infty}^2$, where we have taken the mass of each star to be $1$~M$_\odot$ \citep[e.g.,][equation 3.52]{BinneyTremaine}. As $r_{p}$ increases, we approach the limit where the two stars barely graze, and the deflection angle $\Delta \theta$ should approach $\theta_{\rm hyp}$.

The limiting case described above motivates our fitting formula. Specifically, we find that a function with the form:
\begin{eqnarray} \label{eq:theta_fit}
\Delta \theta = \left(1+A e^{-\left(\frac{r_{p}}{R_\odot} \right)^{2}} -B \frac{v_{\infty}}{v_{\rm esc}} e^{-a \left(\frac{r_{p}}{R_\odot} \right)^2} \right) \theta_{\rm hyp} ,
\end{eqnarray}
fits the data well with $A = 0.16$, $B = 0.35$, $a = 2.5$, 
and $\theta_{\rm hyp}$ given by Eq.~\ref{eq:theta}. While Eq.~\ref{eq:theta} is written to depend on the impact parameter at infinity, $b$, it is related to $r_p$ by:
\begin{eqnarray} \label{eq:r_p_to_b}
b^2 = r_p^2 + 2G(M_{1}+M_{2}) r_p/v_{\infty}^2 \ ,
\end{eqnarray}
where once again we have $M_{1} = M_{2} = 1$~M$_\odot$. We show the predictions of this fitting formula compared to the SPH data and $\theta_{\rm hyp}$ in Figure~\ref{fig:deflection_angle} for nine values of $r_p$. The root-mean-squared error of the fitting formula predictions is $0.022$ radians. The fitting formula is accurate to within $4$ degrees for all of our SPH data and within $1$ degree for $94 \%$ of the data. Crucially, this equation reproduces $\theta_{\rm hyp}$ in the limit $r_p \gtrsim 2$~R$_\odot$ because the exponential terms become small. We also find that the hyperbolic deflection angle for point particles (Eq.~\ref{eq:theta}) is accurate to within $0.1$ radians, or $6$ degrees, for $r_p > 0.6$~R$_\odot$. This result suggests that despite experiencing a physical collision, stars in this regime behave far more like point particles than previously assumed in the literature \citep[e.g.,][]{YuTremaine,Rose&Mockler25}. We attribute this behavior to the fact that most of the star's mass is concentrated near its core, and for larger periapses ($r_p>0.6$~R$_\odot$), only the more diffuse envelopes of the stars interact.

We note that for $r_p = 0.2$~R$_\odot$ and $0.4$~R$_\odot$, the deflection angle becomes negative for very high $v_\infty$ (see first and second panels in Figure~\ref{fig:deflection_angle}). Here, we define a positive deflection angle to denote a deflection towards the other star, as would occur in a purely gravitational interaction. 
A negative deflection angle can occur for the high-speed encounters that push right up to, but do not cross, the threshold of complete disruption while still yielding a two-star outcome. Collisions in this regime generate significant shocks and ejecta, and the bound mass that remains within each star recoils from the impact. A negative deflection angle is tantamount to the stars ``bouncing'' off of each other.
We also note that at the other extreme, low speed and larger $r_p$ collisions, Eq.~\ref{eq:theta} -- for point particles -- slightly under-predicts the deflection angle. For an example, see the last SPH data point versus the grey dashed line for $r_p = 1$~R$_\odot$ or $r_p = 1.4$~R$_\odot$ subplots in Figure~\ref{fig:deflection_angle}. Dissipation during these $v_\infty = 100$~km/s collisions means the stars slow down by $\gtrsim 30 \%$ (see Section~\ref{sec:dv_v_fit} below). It is easier to deflect slower moving stars. However, Eq.~\ref{eq:theta} only depends on $v_\infty$. The reduction of the stars' speed during the collision translates to a slight boost in the deflection angle relative to the prediction from Eq.~\ref{eq:theta}.

Recently, \citet{Rose&Mockler25} considered the orbital effects of stellar collisions. Specifically, they propose that a direct collision can place a star on a nearly radial orbit about the SMBH, such that it becomes tidally disrupted while the star retains signatures of the prior collision. They treated two star outcomes using a semianalytic approach based on limiting cases, the same limiting cases met by our fitting formula. For a head-on collision, the deflection angle approaches zero, and for a grazing encounter, we recover the point particle limit. Therefore, the qualitative behavior between \citet{Rose&Mockler25} for high-speed collisions and our results here are similar. However, their approximation underestimates the deflection angle compared to the SPH. As a result, the rate of collision-induced TDEs from \citet{Rose&Mockler25} is likely conservative. Future work will re-examine the astrophysical implications of direct collisions for SMBH transients.

\subsection{Fractional Change in Speed} \label{sec:dv_v_fit}

In hit-and-run collisions, a star can both change direction (Section~\ref{sec:deflection_angle_fit}) and speed. As touched on in Section~\ref{sec:f_ML_fit}, there are two ways of dissipating kinetic energy from the system: tidal dissipation and shock dissipation. The change in kinetic energy due to tidal dissipation can be described as:
\begin{eqnarray} \label{eq:deltaE_tides}
\Delta E_{\rm tidal} =  \left( \frac{G M_{1}}{R_{1}} \right)^2 \left(\frac{M_{2}}{M_{1}}\right)^2 \sum_{l=2,3...} \left( \frac{R_{1}}{r_p}\right)^{2l+2}T_l(\eta)\ ,
\end{eqnarray}
where $l$ is the spherical harmonic and $T_l(\eta)$ are dimensionless functions of $\eta = \left(\frac{M_{1}}{M_{1} + M_{2}}\right)^{0.5} \left( \frac{r_p}{R_{1}} \right)^{1.5}$ that quantify the deposition of the kinetic energy into the star's internal oscillations \citep[e.g.,][]{PressTeukolsky77,Lee&Ostriker86}.
For our collisions, we can adapt the above equation to:
\begin{eqnarray} \label{eq:deltaE_tides_onlyT2T3}
\Delta E_{\rm tidal} = \frac{G  M_\odot^2 R_\odot^5}{r_p^6} T_2(\eta)+\frac{G  M_\odot^2 R_\odot^7}{r_p^8} T_3(\eta).
\end{eqnarray}
While in principle this calculation involves summing over all spherical harmonics, the other terms prove to be insignificant compared to quadrupole and octupole tides \citep[e.g.,][]{Lee&Ostriker86}. From $\Delta E$, the change in the star's speed can be found using:
\begin{eqnarray} \label{eq:deltav_tides}
\Delta v_{\rm tidal} = \sqrt{E_i/\mu} - \sqrt{(E_i-\Delta E_{\rm tidal})/\mu},
\end{eqnarray}
where $E_i = \frac{1}{2} \mu v_\infty^2$ and $\mu$ is the reduced mass. In the limit of low-speed grazing collisions, we should recover a change in speed given by Eq.~\ref{eq:deltav_tides}. 

\citet{Mardling+01} extend this framework for tidal dissipation to hyperbolic encounters by substituting $\zeta$ for $\eta$. They define $\zeta = \eta \sqrt{2/(1+e)}$, where $e$ is the eccentricity of the orbit between the two stars. This equation recovers $\eta$ for parabolic encounters, while also accounting decreased tidal dissipation at large $v_\infty$.
We opt to use $\zeta$ in place of $\eta$ when predicting the change in speed for our hyperbolic collisions using the fitting formula below.
However, we note that both $\eta$ and $\zeta$ are less than $2$ for a physical collision.

We treat the change in speed of the stars by breaking it into two components, a tidal contribution and a collision contribution:
\begin{eqnarray} \label{eq:dv_v_general}
\frac {\Delta v}{v_\infty}= \left(\frac{\Delta v}{v_\infty} \right)_{\rm tidal}+\left(\frac{\Delta v}{v_\infty} \right)_{\rm coll}. 
\end{eqnarray}
For grazing encounters, $\left(\frac{\Delta v}{v_\infty} \right)_{\rm tidal}$ should dominate and the collision term should approach zero. Furthermore, for close passages that are not physical collisions, our fitting formula should recover the expectation from \citet{PressTeukolsky77} and \citet{Lee&Ostriker86}.

We begin by fitting $\Delta v/v_\infty$ from our SPH dataset for cases with $r_p>1 R_{\odot}$ with a tidal term given by combining Eq.~\ref{eq:deltaE_tides_onlyT2T3} and \ref{eq:deltav_tides}. In this regime, the mass loss is negligible (see Figure~\ref{fig:fractional_mass_loss_fits}) and $\mu$ can be taken as constant. Specifically, we fit the values of $T_2$ and $T_3$ to reproduce our data. To ground our fit in the limiting case of a close encounter, we calculate the actual values of $T_2$ and $T_3$ based on our star's MESA model using GYRE for the relevant eigenfrequencies and eigenfunctions \citep[for GYRE, see][]{Townsend&Teitler13,Townswend+18,Gondstein&Townsend2020,Sun+23}. The above calculation yields similar values to $T_2$ and $T_3$ for an $n = 3$ polytrope as shown in \citet{Lee&Ostriker86}. We ensure that as $r_p$ becomes $>2$~R$_\odot$, our fits for $T_2$ and $T_3$ converge on the GYRE data. 

\begin{figure*}
	\includegraphics[width=0.99\textwidth]{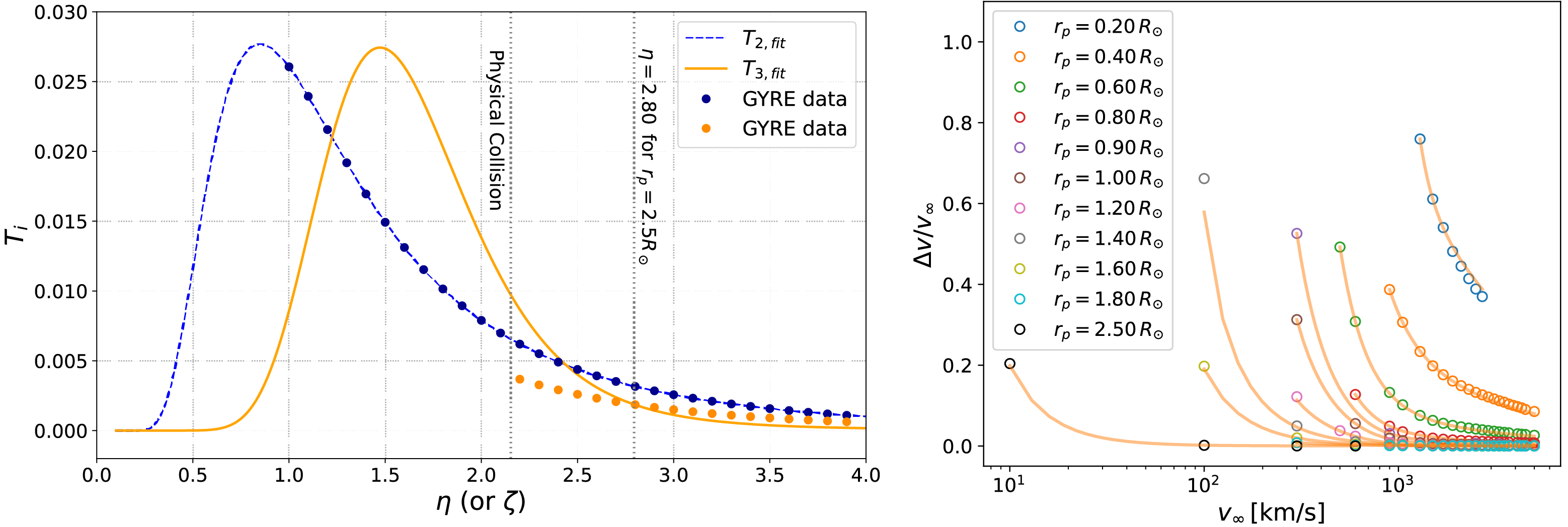}
	\caption{On the left, we show our $T_{2,\mathrm{fit}}$ and $T_{3,\mathrm{fit}}$ for Eq.~\ref{eq:deltaE_tides_onlyT2T3}, which are dimensionless functions that quantify a star's internal response to a tidal field. For $r_p>2 R_{\odot}$, our fits begin to converge with $T_2$ and $T_3$ as calculated for tidal capture using GYRE. On the right, we compare our fitting formula for the change in speed with the SPH data for hit-and-run collisions. The orange curves show our fitting formula, while the open circles represent the SPH data.}
     \label{fig:speed_change}
\end{figure*}

We fit both $T_2$ and $T_3$ with sixth order polynomials in terms of a dimensionless number $\eta$ (or $\zeta$). In order of highest to lowest order term, $T_{2,\mathrm{fit}}(\eta)$ has the coefficients:
\begin{eqnarray*}
[ 0.3082, -3.091,  4.719, -4.510, -0.7109, -1.584]
\end{eqnarray*}
while $T_{3,\mathrm{fit}}(\eta)$ has the coefficients:
\begin{eqnarray*}
[  4.246, 9.556, 7.428, -4.699, -17.09, 5.967, -2.067].
\end{eqnarray*}
We plot $T_{2,\mathrm{fit}}(\eta)$ and $T_{3,\mathrm{fit}}(\eta)$ as well as the GYRE values on the left in Figure~\ref{fig:speed_change}. We confirm that for $T_{2,\mathrm{fit}}$ and $T_{3,\mathrm{fit}}$, 
the tidal term in Eq.~\ref{eq:dv_v_general} becomes negligible for $r_p < 1$~R$_\odot$.

We took cues for the shape of $\left( \Delta v/v_\infty \right)_{\rm coll}$ from the SPH data for $r_p = 0.2$~R$_\odot$, where the collision term should dominate. We find that the collision contribution to Eq.~\ref{eq:dv_v_general} can be fit using a sum of power-laws:
\begin{eqnarray} \label{eq:dv_v_coll}
\left(\frac{\Delta v}{v_\infty} \right)_{\rm coll} = (10^A) \left(\frac{v}{v_\infty} \right)^{-a}+(10^B) \left( \frac{v}{v_\infty} \right)^{-b}. 
\end{eqnarray}
$A$ can be computed using a sixth order polynomial in terms of $\frac{r_p}{R_\odot}$ with coefficients:
\begin{eqnarray*}
[-53.03, 130.5, -64.13, -91.30, 121.8, -53.93, 8.145],
\end{eqnarray*}
while $\log_{10}{a}$ can also be represented by a sixth order polynomial in terms of $\frac{r_p}{R_\odot}$ with coefficients:
\begin{eqnarray*}
[-18.60,  46.79, -22.74, -33.37, 42.43, -17.09, 2.865].
\end{eqnarray*}
The second power-law in Eq.~\ref{eq:dv_v_coll} has $b =0.632$ and $B = 0.874 - 2.596 \frac{r_p}{R_\odot}$.

We show our full fitting formula, Eq.~\ref{eq:dv_v_general}, versus the SPH data on the right in Figure~\ref{fig:speed_change}. The maximum error for the fit is $\lesssim 10\%$. In fact, there is a single data point at $r_p = 1.4$~R$_\odot$ and $v_\infty = 100$~km/s which is under-predicted by more than a few percent by the fitting formula. Otherwise, the fitting formula predicts the change in speed within $2 \%$ for the SPH data.

\section{Machine Learning} \label{sec:machinelearing}

ML techniques offer tantalizing new prospects for modeling large datasets. In this section, we explore the efficacy of ML models compared to physically-motivated analytic expressions like fitting formulae. In our simple two-dimensional parameter space, we test both the commonly used k-Nearest Neighbors (kNN) algorithm and a neural network (NN). We use the \texttt{SciKitLearn} package for the former and \texttt{PyTorch} for the latter. We train classifiers to predict the collision outcome, where labels $0$, $1$ and $2$ denote the number of remnants. Regression is required to predict continuous variables like the fractional mass loss or deflection angle.

For both ML techniques, the input data includes $\log_{10}{v_\infty}$ and $r_p$. Furthermore, we scale the data so that the values lie between $0$ and $1$ for kNN, and we standard normalize the data for the NN. For the NN, we split the dataset into $60 \%$ training data, $20 \%$ validation data, and $20 \%$ testing data. Validating the model during the training loop helps prevent over-fitting of the training data, while the testing data is used to measure the performance of the ML model on unseen data. When splitting our data into the different subsets, we ensure that each class label is represented proportionate to its overall prevalence in the dataset. In other words, if mergers represent the outcome in roughly $5 \%$ of all of our SPH simulations, they will also represent roughly $5 \%$ of the training, testing, and validation datasets. We similarly split the data into $80 \%$ training data and $20 \%$ testing data for the kNN model. However, for kNN, we combine training and validation into one dataset and perform cross-validation within a grid search, described in more detail below.

\subsection{Classification} \label{sec:classification}

The kNN algorithm is an intuitive and effective approach to modeling data \citep{Cover&Hart67}. It predicts the class of a particular data point based on the class of a specified number $k$ of its closest neighbors. To optimize the algorithm for our dataset, we perform a grid search with five-fold cross validation to find the best hyperparameters for the model. We save the model with the highest balanced accuracy score on the validation set. In the grid search, we consider five options for the number of neighbors ($3, 5, 7, 9,$ and $13$); uniform and inverse-distance weights for those neighbors; Minkowski and Chebyshev metrics; and values of $p=1$ and $p=2$ for the power parameter. We note that the power parameter values are only applicable to the Minkowski metric, with $p=1$ corresponding to the Manhattan distance and $p=2$ corresponding to the Euclidean distance. The Chebyshev metric takes the maximum difference between any of the coordinates, in this case, either along the $r_p$ or $log_{10}(v_\infty)$ axes. The grid search returned the Minkowski metric with $p=1$, $3$ neighbors and uniform weights.



NNs perform a series of transformations and activation functions to map input variables onto target output variables. Our NN has two hidden layers with $64$ and $8$ neurons, each followed by a Rectified Linear Unit (ReLU) activation layer \citep[e.g.,][]{fukushima1969visual,glorot2011deep,maas2013rectifier,Agarap2018}. We set the learning rate to $0.1$ and train the NN over $5000$ epochs.\footnote{We also tried learning rates of $0.01$ and $0.001$ and found that they yielded similar performance so long as we increased the number of epochs; the loss rate simply took longer to plateau.} We use Stochastic Gradient Descent as our optimization algorithm \citep[e.g.,][]{robbins1951stochastic,Rumelhart+86} with a cross entropy loss function and verify throughout the training loop that both the training and validation losses decrease.

\begin{figure*}
\begin{center}
	\includegraphics[width=0.99\textwidth]{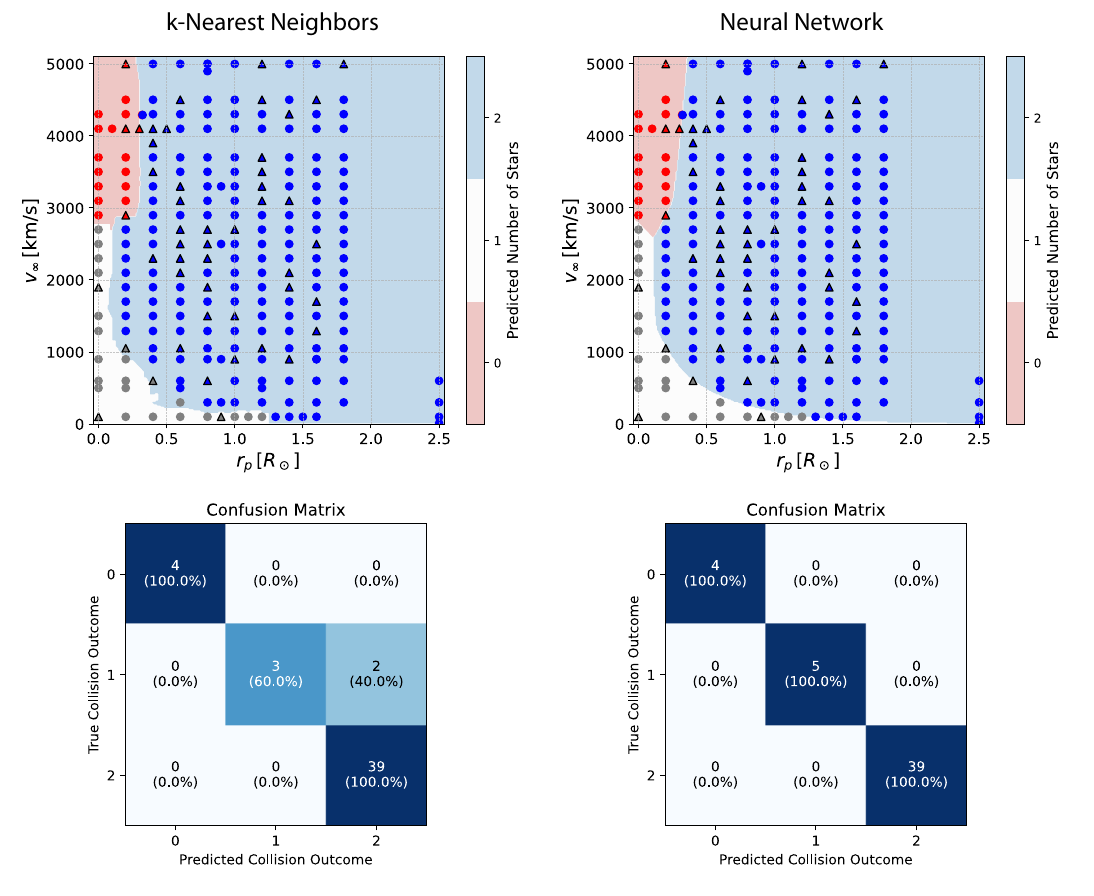}
	\caption{We juxtapose the results of the kNN algorithm, on the left, with our NN, on the right, for classification. In the first row, we show the grid of SPH data plotted over the decision boundaries predicted by the ML models. Circles represent the datasets used for training the ML models, while triangles represent the testing dataset. Grey denotes a merger, blue a two-star outcome (hit-and-run collision), and red a zero-star outcome (total destruction). The color scheme is similar for the decision boundaries, though we use white for one-star outcomes for readability. In the bottom plots, we show the confusion matrix for each ML technique, which shows the ML model's performance across each class of outcome. 
    The percentages in each square indicate the fraction of the data that were correctly or incorrectly classified for each true number of remnants. Because the NN model has $100\%$ accuracy, the off-diagonal terms in its confusion matrix are zero. The NN out-performed kNN. The misclassified data from the latter model always lie near the decision boundaries.}
     \label{fig:ML_classification}
\end{center}
\end{figure*}

We juxtapose our best performing kNN model and NN in Figure~\ref{fig:ML_classification}. The first row shows our SPH data plotted over the decision boundaries of our models. For kNN, the misclassified testing data fall near the model's decision boundary, separating the one and two star collision outcomes. We show the confusion matrices in the second row of Figure~\ref{fig:ML_classification}. As can be seen from the figure, the NN out-performs our kNN model. The accuracy of our best kNN model on the testing data was $95.8 \%$. However, the balanced accuracy is a better measure of the model's performance when one outcome is over-represented in the dataset, as is the case for hit-and-run collisions in our grid. This metric is calculated by averaging the prediction accuracies of the model for each class. The kNN model has a balanced accuracy score of $86.7 \%$ on the testing data. Our NN model, in contrast, has a balanced accuracy of $100 \%$ on the testing data.

\begin{table*}
\caption{Hyperparameters for the best-performing kNN model for each target variable.}
\small
\begin{tabular}{l l c c c c c c}
\hline
\textbf{Mode} & \textbf{Target Variable} & \textbf{k} & \textbf{Metric} & \textbf{Weights} & \textbf{p} \\
\hline
Classification & Number of remnants & $3$  & Minkowski  & Uniform & $1$  \\
Regression  & Fractional mass loss ($f_{\rm ML}$)  & $5$  & Chebyshev  & Inverse distance & NA \\
Regression  & Deflection angle ($\Delta \theta$) & $3$ & Minkowski & Inverse distance & $2$  \\
Regression  & Fractional speed change ($\Delta v/v_\infty$) & $3$ & Chebyshev & Inverse distance & NA  \\
\hline
\end{tabular}
\begin{tablenotes}
\small
 \item $k$ indicates the number of neighbors. 
 \item The hyperparameter $p$ is not applicable (NA) for the Chebyshev metric.
\end{tablenotes}
\label{tab:knn_hyperparams}
\end{table*}

\begin{table*}
\centering
\caption{Performance of different methods for predicting collision outcomes and properties.}
\small
\begin{tabular}{l l c c c c c c}
\hline
\textbf{Method} & \textbf{RMSE} ($f_{\rm ML}$) & \textbf{MAE} ($f_{\rm ML}$) & \textbf{RMSE} ($\Delta \theta$) & \textbf{MAE} ($\Delta \theta$) & \textbf{RMSE} $\left( \frac {\Delta v}{v_\infty} \right)$ & \textbf{MAE} $\left( \frac {\Delta v}{v_\infty} \right)$ \\
\hline
k-Nearest Neighbors  & $0.013$ & $0.07$ & $0.066$ rad & $0.21$ rad & $0.034$ & $0.21$ \\
Neural Network  & $0.008$ & $0.03$  & $0.008$ rad & $0.04$ rad & $0.012$ &  $0.07$ \\
Fitting Formulae  & $0.007$  & $0.04$  & $0.02$ rad & $0.06$ rad & $0.007$ &  $0.08$ \\
\hline
\end{tabular}
\begin{tablenotes}
\small
 \item We indicate the RMSE and maximum absolute error (MAE) for each method and target variables. While these metrics are unit-less for the fractional mass loss and change of speed, they have units radians for the deflection angle.
\end{tablenotes}
\label{tab:accuracy}
\end{table*}

\subsection{Regression} \label{sec:regression}

Similar to our approach to classification, we find the best kNN model for regression using a grid search. The hyperparameters considered are the same as in Section~\ref{sec:classification}. We repeat the grid search for each target variable, $f_{\rm ML}$, $\Delta \theta$, and $\Delta v/v_\infty$, in case they each benefit from different hyperparameters and use the kNN model with the lowest mean squared error (MSE). The hyperparameters returned for each target variable are listed in Table~\ref{tab:knn_hyperparams}. We note that all regression models showed preference for inverse-distance weighting. 

We also use a NN to predict the mass loss and velocity effects of the collision. Our NN has three layers with $128$, $64$ and $32$ neurons each, and uses ReLU as activation functions. We use the MSE as the loss function. We train the NN over $2000$ epochs with a learning rate of $0.01$ and an Adaptive Moment Estimation (Adam) optimization algorithm \citep{Kingma&Ba14,LoshchilovHunter17}.

We evaluate and compare these two methods using the root MSE (RMSE) of their predictions for the testing data. The RMSE for $f_{\rm ML}$, $\Delta \theta$, and $\Delta v/v_\infty$ for kNN were $0.041$ (unit-less), $0.066$ radians, and $0.034$ (unit-less), respectively. The NN delivered lower RMSE values across the board, at $0.008$, $0.008$ radians, and $0.012$
for the target variables in the order listed above. We list these values in Table~\ref{tab:accuracy} for easy comparison. However, in physical systems, we also must care about the absolute errors and how they vary over the parameter space: the RMSE for a predictive model may be low, but it can still be inaccurate in regimes where the deflection angle or fractional mass loss are significant. 

\begin{figure*}
\begin{center}
	\includegraphics[width=0.90\textwidth]{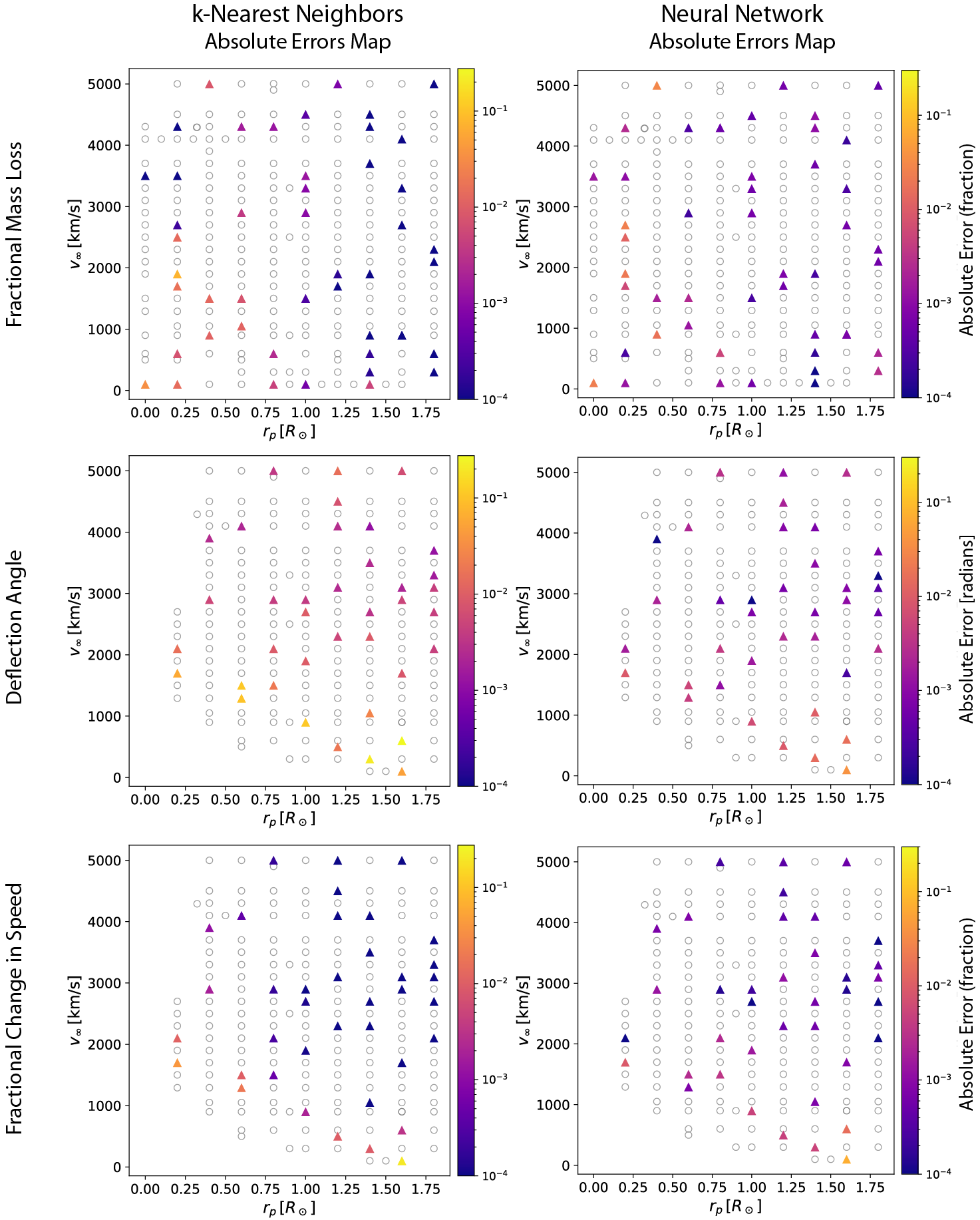}
	\caption{We show a grid of the ML regression errors in our $v_\infty$ versus $r_p$ parameter space, similar in layout to those in Figure~\ref{fig:simulation_grids}. Per our convention in other figures, circles represent training data, and triangles represent testing data. The triangles are color-coded by the absolute error of the prediction compared to the true value. Each row corresponds to one of our target variables. We stress that the absolute error for the fractional mass loss and fractional change in speed is unit-less and can be understood as a fractional error, or a percent error if multiplied by $100$. However, the absolute error in deflection angle has units radians. The kNN model produced larger absolute errors compared to the NN. For both methods, the absolute errors were larger near the boundaries between classes, for example where the outcome transitions from mergers to a two star outcome.}
     \label{fig:NN_vs_kNN_regression}
\end{center}
\end{figure*}

Figure~\ref{fig:NN_vs_kNN_regression} displays the absolute errors of the kNN models and the NN for the testing data over the parameter space. As is our convention, circles represent the training and validation data, while triangles represent the testing data. We color-code the testing data according to the absolute error of the prediction compared to the true value. Consistent with our findings for classification and the regression MSEs, the NN has smaller absolute errors compared to kNN. We list the maximum absolute error for each target variable in Table~\ref{tab:accuracy}. For example, kNN predicts the deflection angle to within $0.2$ radians, while the NN is accurate to within $0.04$
radians. However, both models have the largest absolute errors near boundaries between the collision outcome classes. The regions in the bottom left quadrants of the plots, near the boundary between mergers and hit-and-run collisions, are also where variables like the deflection angle and fractional change in speed exhibit the greatest variation and range in values. Both models tend to be very accurate in the upper right quadrant of the parameter space, where the fractional mass loss, deflection angle, or fractional change in speed are all roughly zero
(see Figure~\ref{fig:simulation_grids}).





\subsection{Comparing Machine Learning and Fitting Formulae} \label{sec:ML_vs_FF}

ML and fitting formulae both have merits when it comes to modeling large sets of data. As analytic expressions, fitting formulae make it easy to check limiting cases and facilitate a physical understanding of the system. However, while we have chosen to focus on sampling a two-dimensional parameter space ($r_p$ and $v_\infty$) in detail, the outcome of a collision also depends on the mass ratio and type of the colliding objects, their structures, and ages. In dense stellar environments, the types of collisions are as diverse as the stellar and compact object populations that reside there. Depending on the behavior of the variable being fitted and the vastness of the parameter space, fitting formulae can become increasingly cumbersome to formulate and to implement. ML therefore represents a promising alternative for modeling collisions in dense stellar environments, and forthcoming work by Gonz\'{a}lez Prieto et al. (in prep) is performing a detailed comparative analysis of various ML algorithms across a broader parameter space.  

In this section, we assess whether ML can out-perform analytic equations which have been expressly tailored to our SPH dataset. As the NN proved more accurate than our kNN model, we use it for this comparison. The RMSE for our fitting formulae predictions compared to the SPH data were $0.007$ and $0.009$ for the fractional mass loss from the mergers and hit-and-run collisions (Eq.~\ref{eq:fML_mergers} and \ref{eq:fML_highspeed}, respectively); $0.02$ radians for the deflection angle (Eq.~\ref{eq:theta_fit}); and $0.007$ for the fractional speed change (Eq.~\ref{eq:dv_v_general}). These values are included in Table~\ref{tab:accuracy} for comparison to the ML methods, though in the table, we calculate a RMSE for both $f_{\rm ML}$ fitting formulae predictions combined.

\begin{figure}
\begin{centering}
	\includegraphics[width=0.75\columnwidth]{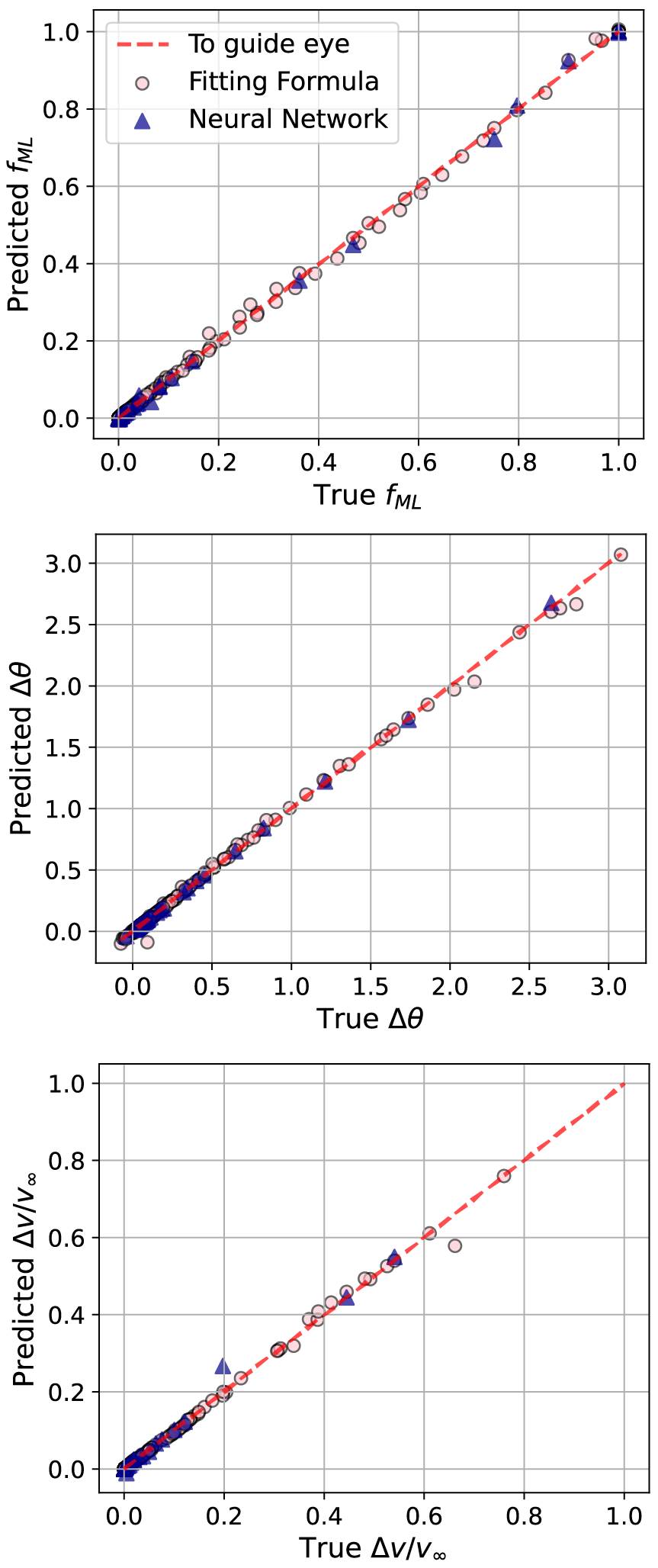}
	\caption{We juxtapose the predictions of our NN with those of the fitting formula. The NN predictions for the testing data versus the true values are shown in dark blue triangles. The pink circles represent the fitting formula predictions versus their true values. We also include the dashed red line to guide the eye. The closer to this line the points fall, the more accurate the prediction. The absolute error of the prediction is given by the vertical distance between the point and the line. For all of our target variables, the NN and fitting formulae have comparable accuracy.}
     \label{fig:NN_vs_FF}
     \end{centering}
\end{figure}

Furthermore, we use Figure~\ref{fig:NN_vs_FF} to visualize their relative performances for the fractional mass loss (first row), deflection angle (second row), and fractional change in speed (third row). The pink dots represent the predicted value of each target variable from the fitting formulae versus their true value. The blue triangles represent the same for the testing dataset from the NN. To guide the eye, we include a dashed red line indicating where the prediction equals the true value. The absolute error of the prediction is given by the vertical distance between the point and the dashed line. While not shown to avoid over-crowding Figure~\ref{fig:NN_vs_FF}, the NN predictions for the training and validation data exhibit a similar high degree of accuracy. 

We find the performance of the NN to be on par and in some cases exceeding the fitting formulae. For all target variables, the maximum absolute error from the NN was less than or equal to the maximum absolute error from the fitting formulae. For the fractional change in speed, both the fitting formula and the NN have a large absolute error (under $10 \%$) for a single data point. What these two data points have in common is that they both lie very near the capture boundary between one and two star outcomes: at $r_p = 1.6$~R$_\odot$ and $v_\infty = 100$~km/s for the NN and $r_p = 1.4$~R$_\odot$ and $v_\infty = 100$~km/s for the fitting formula. The accuracy of the fitting formulae and NN for the fractional mass loss are nearly identical, and the NN predicts the deflection angle with greater accuracy at low $v_\infty$ (see discussion at the end of Section~\ref{sec:deflection_angle_fit}) compared to both Eq.~\ref{eq:theta_fit}, the fitting formulae, and Eq.~\ref{eq:theta}, the deflection angle for point particles experiencing a hyperbolic encounter.


\section{Conclusion} \label{sec:conclusion}

Stellar collisions can occur in dense environments like nuclear star clusters, where they can lead to the formation of unusual stellar populations and drive the production of transient events. We explore a two-dimensional parameter space in $r_p$ and $v_\infty$ for equal mass collisions between $1$~M$_\odot$ stars using the SPH code \texttt{StarSmasher}. We sample initial relative speeds between the colliding stars ranging from $100$ to $5000$~km/s, encapsulating the velocities expected within the sphere of influence of a SMBH. We sample the distance of closest approach, or periapsis $r_p$, of the encounter between $0$ and $1.8$~R$_\odot$ for a physical collision. We supplement our dataset of physical collisions with four simulations of close passages, $r_p = 2.5$~R$_\odot$, to understand how the kinematics of collisions relate to expectations from tidal dissipation.

We begin by categorizing the collisions by qualitative outcome, which can be classified according to the number of collision remnants. In the first case, the stars are able to capture each other and merge, producing a single star. In the second case, the relative speed between the stars is either too high or the periapsis too large for the stars to capture one another and merge. We refer to these cases as hit-and-run type collisions, and they result in two collision remnants. Lastly, the collision can completely destroy the stars, leaving behind zero collision remnants. These collisions typically have large $v_\infty$ and small $r_p$.

We use our two-dimensional grid to build a physical intuition about the effects of stellar collisions and develop fitting formulae for three key properties of the collision remnants. The first property is $f_{\rm ML}$, the fractional mass loss, which quantifies the mass ejected from the system during the collision. The second two quantities relate to the kinematics of the collision remnants. In collisions with symmetric mass loss, as is the case for our equal-mass collisions, the merger product remains at rest in the center of mass frame. In hit-and-run collisions; however, the stars continue along some modified trajectory. We understand changes to the stars' velocity vectors in two parts: the change in speed and the deflection angle. Lastly, we use our grid to test two ML algorithms, kNN and a NN, and we compare the accuracy of our best ML model to the fitting formulae. We summarize our results below:

\begin{enumerate}
    \item \textbf{Capture Boundary:} The capture boundary separates mergers from hit-and-run collisions. For a given value of $v_\infty$, there is some critical $r_p$ within which the stars must pass to dissipate their relative kinetic energy and become a bound system. We find that the fitting formulae from \citet{Lai+93} accurately predicts the capture radius for our simulation grid after a simple scaling of their $A_{\rm cap}$ fitted constant to make it applicable to $1$~M$_\odot$ stars. The formula can be found in Eq.~\ref{eq:capture_fit_ourunits}. All physical collisions that produce a bound system ultimately result in a merger \citep[consistent with previous studies, see, e.g.,][]{BenzHills87,Lai+93}. While some low-speed grazing collisions can temporarily produce a binary, the stars in these systems will experience repeated collisions until they coalesce. We note that the boundary transition to fully destructive collisions is given by the fitting formulae for the fractional mass loss, below, when it becomes equal to $1$.
    \item \textbf{Fractional Mass Loss:} The fractional mass loss from the system is defined as $\frac{\Delta M}{M_1+M_2}$. As described above, our collisions have $M_1 = M_2 = 1$~M$_\odot$. Due to symmetry arguments, fractional mass loss from the system is equivalent to the fractional mass loss from each of the stars. We fit the $f_{\rm ML}$ for one and two star collision outcomes separately, though we begin with an equation of the form $f_{\rm ML} = A + B\left( \frac{v_\infty}{v_{\rm esc}} \right)^a$ for both. For nearly head-on collisions, $a$ is approximately $2$. The full fitting formulae are given by Eq.~\ref{eq:fML_mergers} and \ref{eq:fML_highspeed}. The fit for mergers, Eq.~\ref{eq:fML_mergers}, reproduces the expected mass loss for parabolic collisions ($v_\infty \ll v_{\rm esc}$) from \citet{Lombardi+02}. The fitting formulae are accurate to with $4 \%$ for all of our SPH data points.
    \item \textbf{Deflection Angle:} Hyperbolic encounters between point particles informs our fitting formula for the deflection angle of the stars following a hit-and-run collision. Our fitting formula is given by Eq.~\ref{eq:theta_fit}, accurate to within $4$ degrees for all of our SPH data. Furthermore, we find that the deflection angle is well-approximated by the point particle limit (Eq.~\ref{eq:theta}) for $r_p>0.6$~R$_\odot$ because most of a star's mass is concentrated at its center. Generalizing this behavior to all main-sequence stellar collisions, we expect the hyperbolic encounter limit, Eq.~\ref{eq:theta}, to be accurate to within $\sim 6$ degrees for $r_p \gtrsim 0.3 \times (R_1+R_2)$, where $R_1$ and $R_2$ are the radii of the colliding stars.
    \item \textbf{Change in Speed:} In addition to the deflection angle, we consider the fractional change in speed, $\Delta v/v_\infty$, of the stars following a hit-and-run collision. For low-speed grazing collisions, tidal dissipation should be the main contributor to changes in a star's speed. As $r_p$ begins to decrease, shock dissipation should become the dominant means of slowing down the stars. We therefore fit $\Delta v/v_\infty$ using two terms, one representing the collision and one representing the tidal contribution. For large $r_p$, the latter term dominates. As $r_p$ becomes small, the tidal term becomes negligible and the collision term begins to dominate. Section~\ref{sec:dv_v_fit} details our approach and the final fitting formula. The formula is generally accurate to within a couple percent, with the exception of one data point near the capture boundary where $\Delta v/v_\infty$ is underpredicted by $\sim 10 \%$.
    
    \item \textbf{ML Algorithms:} We use our two-dimensional grid as a testing ground for two ML algorithms, kNN and a NN, which offer tantalizing prospects for modeling large datasets. We use classification to predict the number of stellar remnants in Section~\ref{sec:classification} and regression to predict continuous target variables like the $f_{\rm ML}$ in Section~\ref{sec:regression}. We find that the NN outperforms our best kNN model in all respects. The NN classifier correctly predicted the collision outcome for the entire test dataset, while kNN struggled with data points that fell near decision boundaries, which separate one class of collision outcomes from the others. The NN also yielded lower MSEs and maximum absolute errors for all target variables, and both the NN and kNN had the largest absolute errors in their predictions near boundaries between collision outcome classes, where target variables also have the largest variance. Lastly, we found the NN to be on par with and in some cases exceeding the accuracy of fitting formulae, which were expressly tailored for the SPH data at hand.
\end{enumerate}

Fitting formulae and ML both have advantages for modeling datasets. Fitting formulae can provide insights into the physical laws and limits of a system. However, as the parameter space for the initial conditions expands, they can also become increasingly complex to formulate and cumbersome to implement. Our results highlight the promise of NNs to model collisions in complex environments, where the types of collisions are as diverse as the stellar and compact object populations that reside there.

\begin{acknowledgments}
We thank Tjitske Starkenburg and Enrico Ramirez-Ruiz for valuable discussion. S.C.R. thanks the CIERA Lindheimer Fellowship for support. Support for E.G.P.\ was provided by the National Science Foundation Graduate Research Fellowship Program under Grant DGE-2234667. F.K. acknowledges support from the CIERA Postdoctoral Fellowship. We gratefully acknowledge the support of the NSF-Simons AI-Institute for the Sky (SkAI) via grants NSF AST-2421845 and Simons Foundation MPS-AI-00010513. This research was supported in part through the computational resources and staff contributions provided for the Quest high-performance computing facility at Northwestern University, which is jointly supported by the Office of the Provost, the Office for Research, and Northwestern University Information Technology. This work also used Bridges-2 at the Pittsburgh Supercomputing Center through allocation PHY240311 from the Advanced Cyberinfrastructure Coordination Ecosystem: Services \& Support (ACCESS) program, which is supported by U.S. National Science Foundation grants 2138259, 2138286, 2138307, 2137603, and 2138296.

\end{acknowledgments}

\appendix

\section{Resolution Study} \label{sec:resolution_study}

We ran additional simulations to confirm that $10^5$ particles per star were sufficient for convergence of our parameters of interest. We tested five values of the number of particles $N$ per star: namely, $10^4$, $2.5 \times 10^4$, $5 \times 10^4$, $10^5$, and $2 \times 10^5$, and increased the number of neighbors $\propto \sqrt{N}$. We show two examples from this resolution study in Figure~\ref{fig:convergence} of the fractional mass loss obtained versus $N$. We determine that $10^5$ particles per star with $\sim 90$ neighbors is adequate to capture the relevant physics for the purposes of creating a dataset of collision outcomes.


\setcounter{figure}{0}
\renewcommand{\thefigure}{A\arabic{figure}}

\begin{figure}
\begin{center}
	\includegraphics[width=0.85\textwidth]{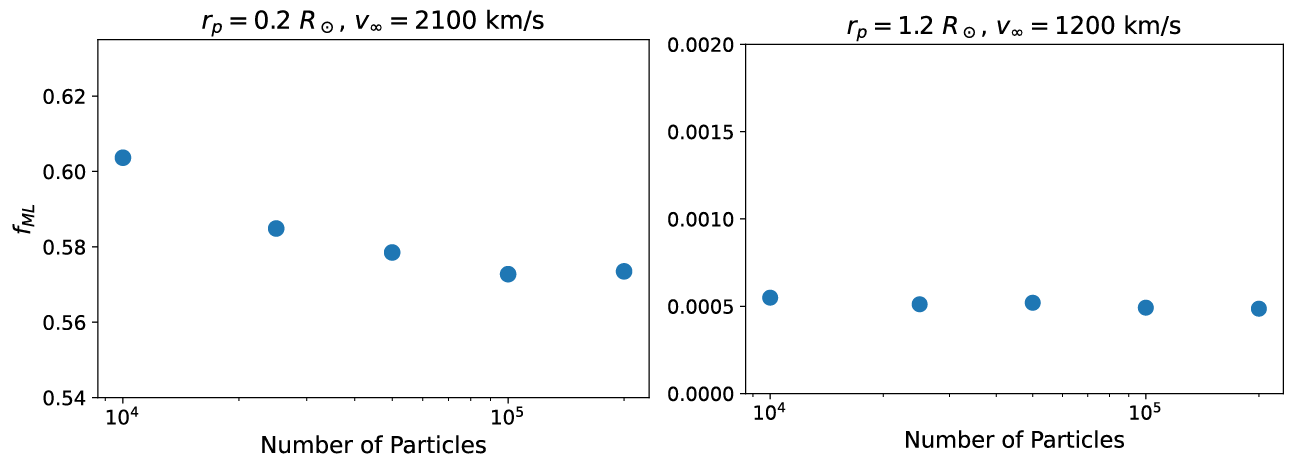}
	\caption{The fractional mass loss for two sets of initial conditions, $r_p = 0.2$~R$_\odot$ and $v_\infty = 2100$ km/s (left) and $r_p = 1.2$~R$_\odot$ and $v_\infty = 1200$ km/s (right) from five simulations of different resolution.}
     \label{fig:convergence}
\end{center}
\end{figure}

\bibliography{sample631}{}
\bibliographystyle{aasjournal}

\end{document}